\documentclass[twocolumn,aps,prl,superscriptaddress
,groupedaddress]{revtex4}
\usepackage{ucs}
\usepackage[english]{babel}
\usepackage{amsfonts}

\usepackage[latin9]{inputenc}
\usepackage{color}
\usepackage{amsmath}
\usepackage{amssymb}
\usepackage{graphicx}
\usepackage{esint}
\usepackage[unicode=true,pdfusetitle,
 bookmarks=true,bookmarksnumbered=false,bookmarksopen=false,
 breaklinks=true,pdfborder={0 0 1},backref=false,colorlinks=true]
 {hyperref}
\hypersetup{
 linkcolor=red,urlcolor=blue,citecolor=red,anchorcolor=blue}
\usepackage{breakurl}

\usepackage{bm}
\usepackage[T1]{fontenc}

\begin{document}
\title{Dimensionally Induced Phase Transition of the Weakly Interacting Ultracold Bose Gas}
\author{Bernhard Irsigler}
\affiliation{Institut f\"ur Theoretische Physik, Johann
Wolfgang Goethe-Universit\"at Frankfurt am Main, Germany}
\affiliation{Institut f\"ur Theoretische Physik, Freie Universit\"at Berlin, Germany}
\author{Axel Pelster}
\affiliation{Fachbereich Physik und Forschungszentrum OPTIMAS, Technische Universit\"at
Kaiserslautern, Germany}
\date{\today}
\begin{abstract}

We investigate the dimensionally induced phase transition from the normal to the Bose-Einstein-condensed phase for a weakly interacting Bose gas in an optical lattice. To this end we make use of the Hartree-Fock-Bogoliubov-Popov theory, where we include numerically exact hopping energies and effective interaction strengths. At first we determine the critical chemical potential, where we find a much better agreement with recent experimental data than a pure Hartree-Fock treatment. This finding is in agreement with the dominant role of quantum fluctuations in lower dimensions, as they are explicitly included in our theory. Furthermore, we determine for the 1D-3D-transition the power-law exponent of the critical temperature for two different non-interacting Bose gas models yielding the same value of 1/2, which indicates that they belong to the same universality class. For the weakly interacting Bose gas we find for both models that this exponent is robust with respect to finite interaction strengths.
\end{abstract}
\maketitle

\section{Introduction}
Low-dimensional systems play an important role in physics, as they can exhibit tremendously different behavior than in the three-dimensional case due to the lack of certain degrees of freedom. Prominent examples are provided by the Tonks-Girardeau gas in 1D  \cite{tonks36,girardeau60} or the Berezinskii-Kosterlitz-Thouless transition in 2D \cite{berezinskii72,kosterlitz73}, which were experimentally observed in the realm of ultracold quantum gases in Ref.~\cite{paredes04} and \cite{hadzibabic06}, respectively. From the Mermin-Wagner-Hohenberg theorem \cite{mermin66,hohenberg67} it is known that there cannot be a one- or two-dimensional, homogeneous Bose-Einstein condensate (BEC) at finite temperatures, see Ref.~\cite{ueda10} for a derivation in the context of ultracold atoms. However, as a three-dimensional BEC does exist, it is expected that an increment of the critical temperature should be observable when going continuously from low to three dimensions. In principle, there are two distinct approaches to induce such a dimensional phase transition of ultracold atomic gases. Low-dimensional systems are commonly achieved by using anisotropic, confining traps \cite{khaykovich02,weller08,kruger10,martin10,dyke11,yefsah13}. Such a dimensional transition concerning Bose gases was investigated theoretically in Refs.~\cite{das02,alkhawaja03,kadio05}. As an example of the dimensionally induced phase transition, the 2D-3D-transition was studied experimentally in Ref.~\cite{dyke11} using an anisotropic harmonic trap.

\begin{figure}[t]
\centering
\includegraphics[trim={0 0 25mm 0},clip,scale=.19]{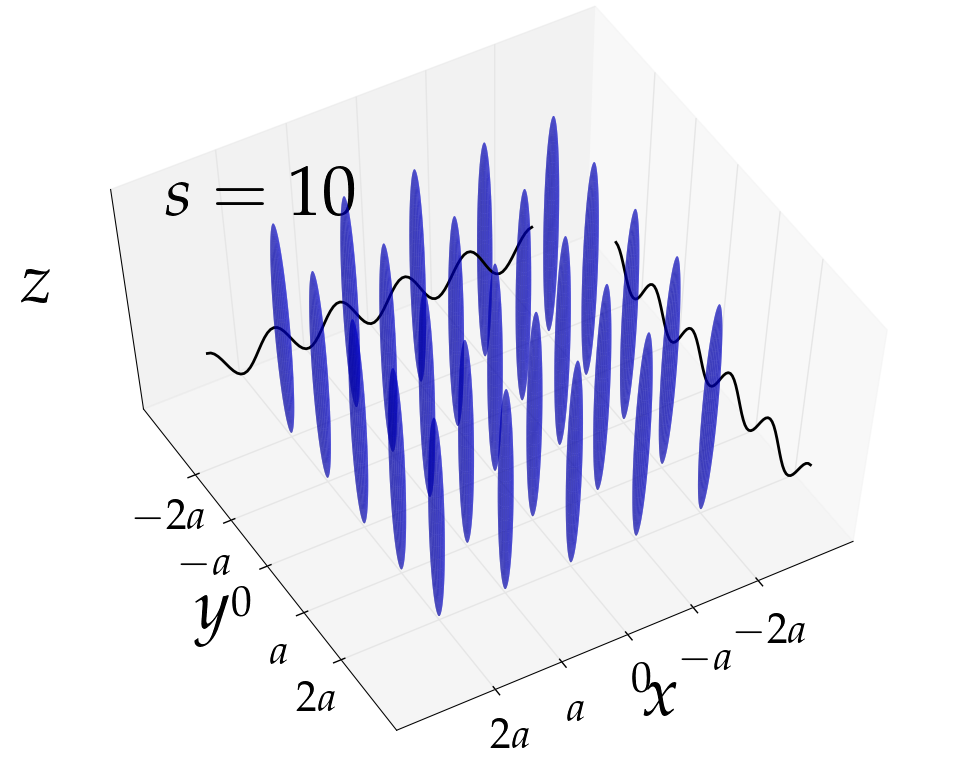}
\includegraphics[trim={0 0 25mm 0},clip,scale=.19]{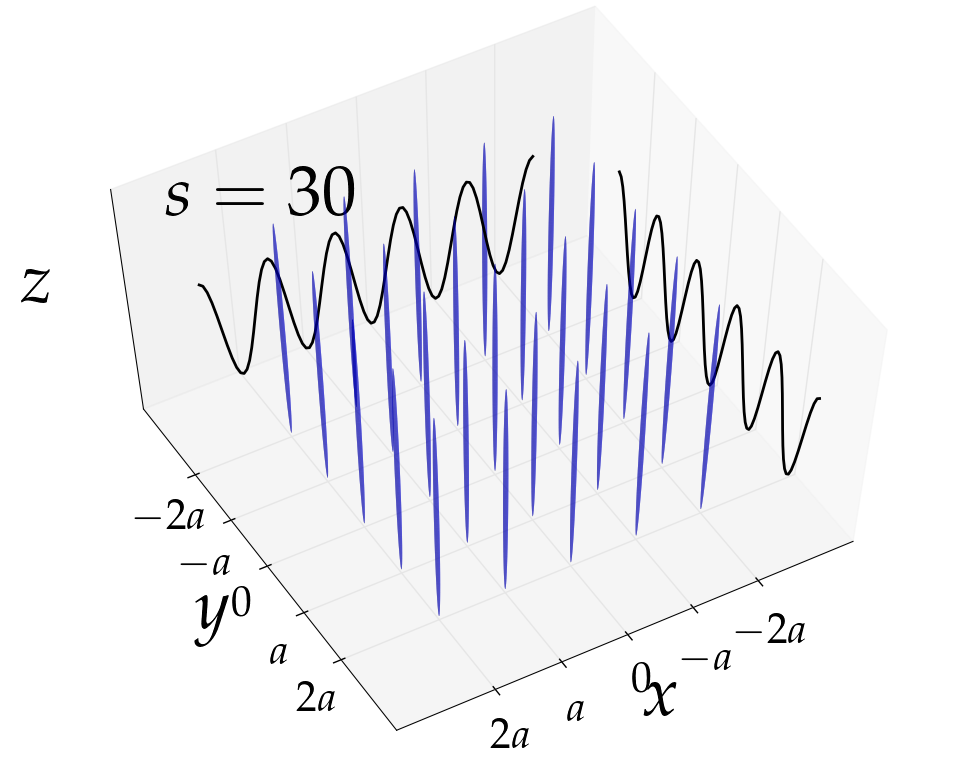}
\caption{Schematic setup of {hybrid model} for two dimensionless lattice depths $s$: bosonic clouds in longitudinal direction are depicted in blue and optical lattice potential in $xy$-plane is represented by black lines.}
\label{fig:schematic}
\end{figure}

Here we follow another approach, which keeps the potential energy unchanged and varies only the hopping energy of a homogeneous Bose gas within an optical lattice. In the following, we investigate at first a {hybrid model} which constitutes a dimensional phase transition. As depicted in 
Fig.~\ref{fig:schematic}, it consists of a three-dimensional optical lattice, without confinement in the longitudinal direction. This system has recently been investigated in an experimental setup \cite{vogler14}. Although experiments are always performed in trapped systems, one can relate them with a corresponding homogeneous case by making use of the local density approximation (LDA). Introducing a local effective chemical potential $\mu_\text{eff}(\bm{r})=\mu-V(\bm{r})$ allows to match local quantities of the inhomogeneous setting with global quantities of the homogeneous theory. The LDA decomposes the actual inhomogeneous problem into one homogeneous problem with the effective chemical potential $\mu_\text{eff}(\bm{r})$ for every value of $\bm{r}$. In the harmonically trapped Bose gas the BEC phase is located in the inner part of the trap, where the overall density is higher. The pure thermal phase stays in the outer region of the trap, where the density is lower. Both phases are spatially separated at the coordinate $\bm{r}_c$, where the density $n(\bm{r}_c)$ equals the critical density $n_c$ for the BEC transition. Using LDA the critical density is thus determined through the homogeneous problem with the corresponding chemical potential $\mu_\text{eff}(\bm{r}_c)$ \cite{vogler14}. This article obtains the critical chemical potential of the inhomogeneous experiment of Ref.~\cite{vogler14} by using a homogeneous theory, which explicitly takes into account the impact of quantum fluctuations.

Within a homogeneous lattice the one-particle dispersion relation consists of band energies. Since we treat ultracold systems, we restrict ourselves to the lowest energy band and nearest neighbor hopping only. The most general form of the dispersion relation of a three-dimensional, orthogonal optical lattice in the tight-binding approximation reads
\begin{equation}
\epsilon_{\bm{k}}=\sum_i2J_i\left[1-\cos(k_ia_i)\right],
\label{dispersionRelation}
\end{equation}
which has been shifted by an energy offset to avoid negative energies. Here $J_i$ and $a_i$ represent the hopping energy and the lattice constant of the spatial dimension $i=x,y,z$, respectively. For $k_ia_i\ll1$ we can approximate Eq.~\eqref{dispersionRelation} by a quadratic expression in $k_i$ such that an effective mass $M^*_i$ can be assigned as
\begin{equation}
\epsilon_{\bm{k}}\approx\sum_i\frac{\hbar^2k_i^2}{2M^*_i},\quad
M^*_i=\frac{\hbar^2}{2J_ia_i^2}.
\label{effectiveMass}
\end{equation}
With this notation the {hybrid model} is characterized by
\begin{equation}
J_x=J_y=J,\quad 
a_x=a_y=a,\quad
J_z=\frac{\hbar^2}{2Ma_z^2},\quad
a_z\rightarrow0.
\label{hybridModel}
\end{equation}
Here $M$ stands for the atomic mass and for the $z$-direction the continuum limit is applied. A schematic setup for the {hybrid model} is presented in Fig.~\ref{fig:schematic} for different lattice depths, where the dimensionless lattice depth is defined as $s=V_0/E_r$. Here, $V_0$ corresponds to the intensity of the laser pair building up the optical lattice and $E_r=\pi^2\hbar^2/(2Ma^2)$ denotes the recoil energy. The dimensional transition is performed by tuning the dimensionless lattice depth $s$. For $s=0$ the lattice is completely ramped down, which corresponds to a pure three-dimensional, homogeneous system. Ramping up the lattice depth causes an emergence of a tube structure as depicted in Fig.~\ref{fig:schematic}. For deep lattices of about $s=30$ the hopping between the tubes is suppressed and the system corresponds to an array of decoupled one-dimensional tubes. Hence, just by varying the lattice depth the 1D-3D-transition can be induced.

This article is structured as follows. In Sec.~\ref{sec2} we briefly review the Hartree-Fock-Bogoliubov-Popov theory, which represents our main formalism, and as a special case the Hartree-Fock theory. In Sec.~\ref{sec4} we explain how the hopping energies as well as the effective interaction strengths behave during the dimensional transition by discussing different approximation methods. In Sec.~\ref{sec5} we move on to the results for the critical chemical potential and compare them with the recent experimental data from Ref.~\cite{vogler14}. In Sec.~\ref{sec6} we present the corresponding findings for the critical temperature in the 1D-3D-transition. Eventually, we conclude with an outlook for further research topics in Sec.~\ref{sec8}.

\section{Hartree-Fock-Bogoliubov-Popov theory}
\label{sec2}
The Hartree-Fock-Bogoliubov-Popov theory (HFBP) \cite{andersen04,pethick08,stoof09,griffin09,pitaevskii16} interpolates between the Bogoliubov theory at zero temperature and the Hartree-Fock theory (HF) at finite temperatures. It treats the weakly interacting bosons approximatively as a gas of non-interacting quasiparticles exhibiting the dispersion relation $E_{\bm{k}}=\sqrt{\varepsilon_{\bm{k}}^2+2gn_0\varepsilon_{\bm{k}}}$ with $\varepsilon_{\bm{k}}=\epsilon_{\bm{k}}-\mu+g(2n-n_0)$. Here $\mu$ denotes the chemical potential in the grand-canonical description, $g=4\pi\hbar^2a_s/M$ represents the three-dimensional interaction strength with $a_s$ being the s-wave scattering length, $n$ stands for the total particle density, and $n_0$ represents the condensate density. The particle density in the HFBP formalism is given by
\begin{equation}
n=n_0+\frac{1}{V}\sum_{\bm{k}}\left[\frac{\varepsilon_{\bm{k}}+gn_0}{E_{\bm{k}}}\left(\frac{1}{e^{\beta E_{\bm{k}}}-1}+\frac{1}{2}\right)-\frac{1}{2}\right],
\label{HFBPdensity}
\end{equation}
where $V$ denotes the volume of the system and $\beta=1/(k_\mathrm{B}T)$. Equation \eqref{HFBPdensity} contains contributions from both the thermal fluctuations of the HF theory and the quantum fluctuations of the Bogoliubov theory. The chemical potential in the condensate phase is given by solving the generalized Gross-Piteavskii equation, which reads for the homogeneous case:
\begin{equation}
\mu=2gn-gn_0.
\label{GPequation}
\end{equation}
For bosons interacting via a two-body contact potential the Hartree and the Fock term coincide, which leads to the first term of the right-hand side of Eq.~\eqref{GPequation}. The second term represents a contribution, which enters through the Bogoliubov channel.  Note that Eq.~\eqref{GPequation} is confirmed by the Hugenholtz-Pines theorem \cite{hugenholtz59,andersen04}, thus HFBP describes a gapless superfluid phase \cite{greiner02}.
At constant total density, Eq.~\eqref{HFBPdensity} yields a first-order phase transition at the critical temperature \citep{andersen04}. Thus, the critical point is determined by a finite critical condensate density $n_{0c}$, where its derivative with respect to the inverse critical temperature $\beta_c=1/(k_BT_c)$ diverges:
\begin{equation}
n_0(\beta_c)=n_{0c},\quad 
\left.\frac{\partial n_0}{\partial\beta}\right|_{\beta_c}=\infty.
\label{firstOrder}
\end{equation}
In order to find the critical point both Eqs. \eqref{firstOrder} have to be solved simultaneously.

Now we briefly review a special case of a pure Hartree-Fock formalism (HF), when the self-energy contribution $gn_0$ of the Bogoliubov channel is not considered. Therefore, the quasiparticle energy reduces to $E_{\bm{k}}=\epsilon_{\bm{k}}+2gn-\mu$ and Eq.~\eqref{HFBPdensity} simplifies to 
\begin{equation}
n=n_0+\frac{1}{V}\sum_{\bm{k}}\frac{1}{e^{\beta(\epsilon_{\bm{k}}+2gn-\mu)}-1}.
\label{HFdensity}
\end{equation}
Here the chemical potential coincides with $2gn$ at the critical point, which is consistent with \eqref{GPequation} if the Bogoliubov contribution is neglected. Note that the HF theory coincides with the non-interacting one since it differs only by a physically irrelevant shift of the chemical potential, i.e., it is independent of $g$. In the thermodynamic limit the sum in Eq.~\eqref{HFdensity} goes over into an integration, which can be performed exactly yielding for the {hybrid model}
\begin{equation}
n=n_0+\frac{1}{a^2\lambda_T}\sum_{m=1}^\infty\frac{1}{m^{1/2}}
e^{m\beta(\mu-2gn)}e^{-4m\beta J}I_0^2(2m\beta J).
\end{equation}
Here $\lambda_T=\sqrt{2\pi\beta\hbar^2/M}$ denotes the thermal de Broglie wavelength and $I_0(x)$ represents the modified Bessel function of first kind \cite[(9.6.16)]{abramowitz64}.

We investigate the critical temperature by setting $n_0=0$ and $\mu=2gn$. Numerical calculations show that $\beta J\rightarrow0$ for $J\rightarrow0$. Thus, the limit $\beta J\rightarrow0$ describes the behavior of the critical temperature deep in the one-dimensional regime. Using the approximation \cite[(9.7.1)]{abramowitz64} for small arguments of the modified Bessel function, we find for the critical temperature $T_c$ of the {hybrid model} as a function of the transverse hopping energy $J$ 
\begin{equation}
\frac{k_BT_c}{E_r}=\frac{4na^3}{\pi}\sqrt{\frac{J}{E_r}}.
\label{HFmodelA}
\end{equation}

In the following we investigate within the HFBP theory, whether a finite two-particle interaction strength changes the above power-law exponent of 1/2. To this end we have to determine how the two energy scales of the Bose-Hubbard Hamiltonian, i.e., the hopping energy and the effective interaction, depend on the dimensionless lattice depth $s$.

\section{Hopping energy and effective interaction}
\label{sec4}
\begin{figure}
\begin{flushleft}
\includegraphics[width=.98\columnwidth]{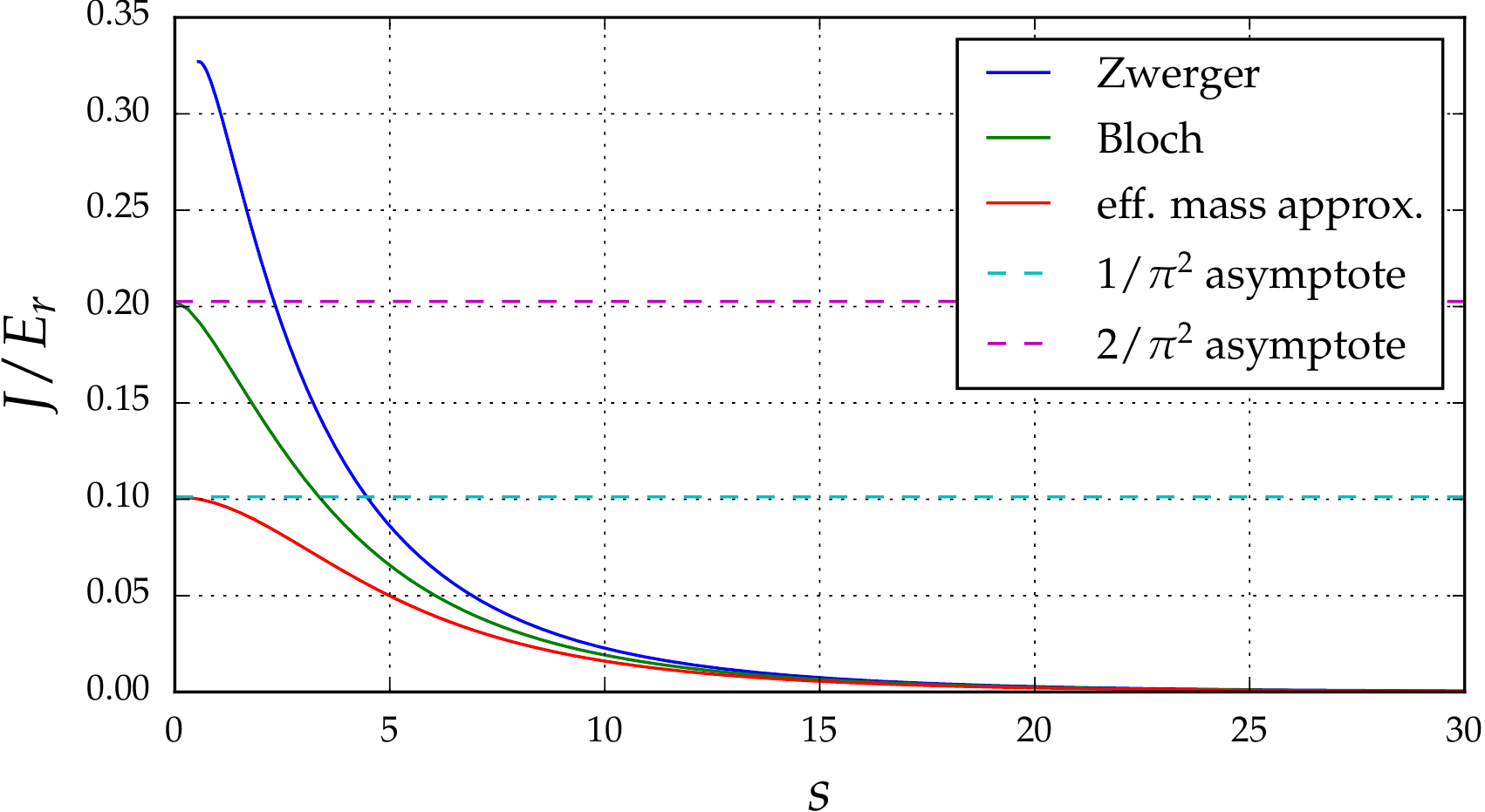}
\caption{Hopping energy $J$  as function of dimensionless lattice depth $s$: solid lines depict from the top to the bottom solutions of three different approximation methods described in the text, respectively. Since the top blue curve defined by Eq.~\eqref{zwerger} vanishes at $s=0$, it is plotted only up to its maximal value.}
\label{fig:hopping}
\end{flushleft}
\end{figure}

Here we discuss three approximation methods to compute the hopping energy $J$ as a function of the dimensionless lattice depth $s$. The first one is an analytic expression of Zwerger \cite{zwerger03}, which is valid for deep optical lattices and follows from approximately solving the one-dimensional Mathieu equation:
\begin{equation}
J=\frac{4}{\sqrt{\pi}}E_rs^{3/4}e^{-2\sqrt{s}}.
\label{zwerger}
\end{equation}
The second one is the numerical solution of the one-dimensional Schr\"odinger equation using the Bloch theorem. In that case the hopping energy follows as the Fourier transform of the band energy dispersion relation \cite{krutitsky16}.
The third method directly follows from the latter dispersion relation by approximating it with a parabolic fit. Thus, an effective mass \eqref{effectiveMass} can be assigned, which itself defines a corresponding hopping energy. We present the respective results for these three methods in Fig.~\ref{fig:hopping}. Therein, we see a good agreement of the three methods for deep lattices. However, for shallow lattices the top blue curve overestimates the hopping energy and becomes non-monotonic below $s\approx1$. Remarkably, the middle green curve and the bottom red curve  differ exactly by  a factor of 2 from the values at $s=0$, i.e., the continuum values \cite{krutitsky16}. This is due to the fact that the relation between $J_i$ and $M^*_i$ in Eq.~\eqref{effectiveMass} stems originally from the tight-binding dispersion \eqref{dispersionRelation}.

\begin{figure}
\begin{flushleft}
\includegraphics[width=.98\columnwidth]{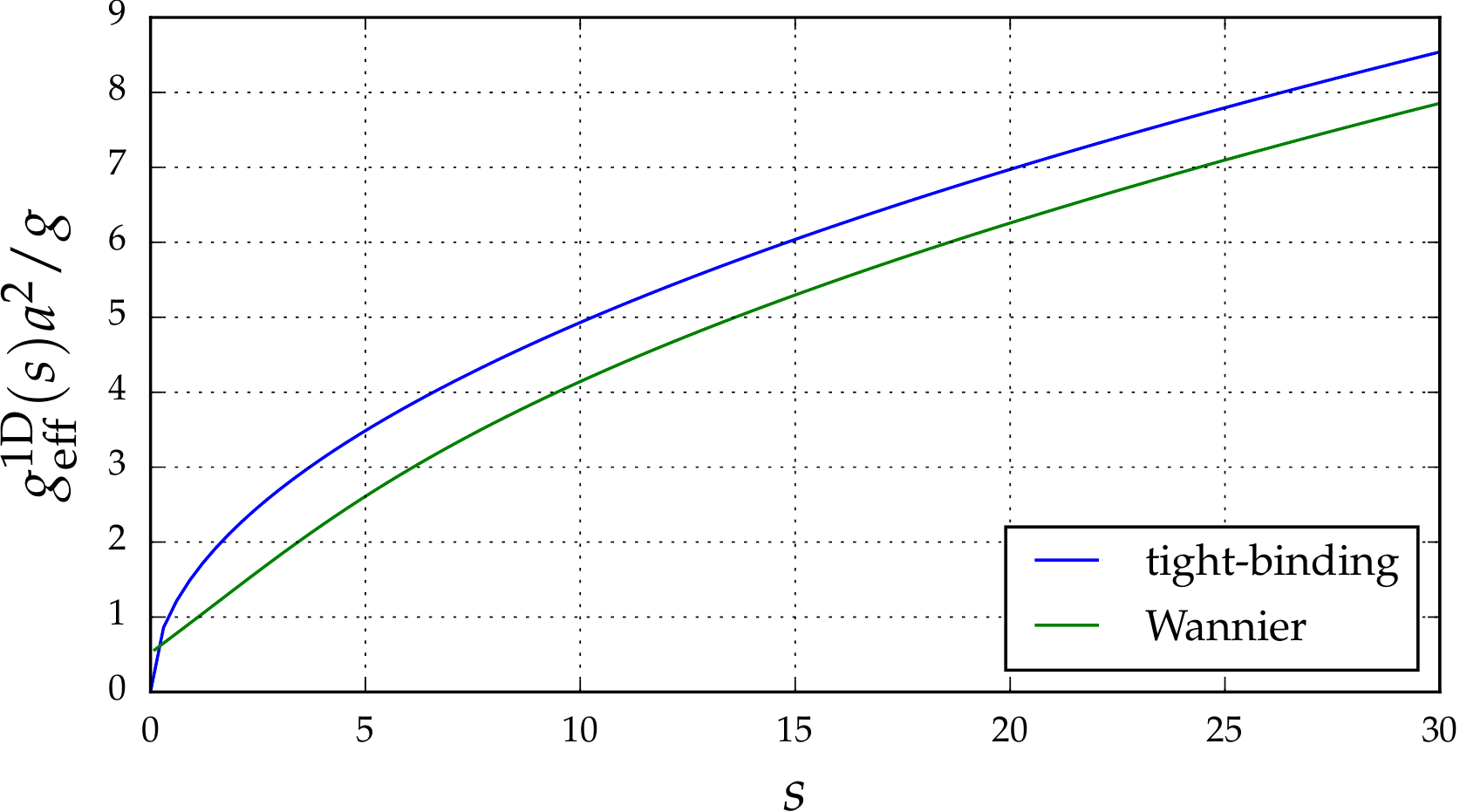}
\caption{One-dimensional effective interaction strength $g_{\mathrm{eff}}^{\mathrm{1D}}$ as function of dimensionless lattice depth $s$. Top curve corresponds to tight-binding approximation \eqref{tightBinding}, lower curve stems from  Eq.~\eqref{interactionWannier} and numerically determined Wannier function.}
\label{fig:interactionStrength}
\end{flushleft}
\end{figure}

Furthermore, besides the hopping energy, also the interaction strength turns out to be a function of the lattice depth. Following the reasoning of Ref.~\cite{vogler14}, we define a one-dimensional, effective interaction strength $g_{\mathrm{eff}}^{\mathrm{1D}}\simeq2\hbar a_s\omega_\perp(s)$ \cite{olshanii98}, where  $\omega_\perp(s)$ represents the transverse trapping frequency of a single tube. Within the tight-binding approximation, $g_{\mathrm{eff}}^{\mathrm{1D}}$ is given by \cite{greiner02}
\begin{equation}
g_{\mathrm{eff}}^{\mathrm{1D}}\approx4a_sE_r\sqrt{s}.
\label{tightBinding}
\end{equation}
We read off that for a vanishing lattice depth the effective interaction strength vanishes, which is not physical, since there must be a finite interaction strength in the pure three-dimensional system. In order to redeem the tight-binding approximation, we calculate the interaction strength also with numerically determined Wannier functions. In the Bose-Hubbard formalism the on-site interaction strength $U$ is given by
\begin{equation}
U=g\int\mathrm{d}\bm{r}|w(\bm{r})|^4,
\label{onSite}
\end{equation}
where $w(\bm{r})$ denotes the Wannier function that factorizes for cubic or quadratic lattices into their respective one-dimensional counterparts. Due to higher coherence with neighboring lattice sites for shallow lattices, the Wannier function delocalizes over the lattice. Thus, the on-site interaction strength decreases with decreasing lattice depth, an effect which is enhanced through the fourth power in Eq.~\eqref{onSite}. Since in the {hybrid model} only two lattice dimensions contribute, the one-dimensional, effective interaction strength is given there as
\begin{equation}
g_{\mathrm{eff}}^{\mathrm{1D}}=g\left[\int\mathrm{d}x|w(x)|^4\right]^2,
\label{interactionWannier}
\end{equation}
where $w(x)$ denotes now the one-dimensional Wannier function. In Fig.~\ref{fig:interactionStrength} we show the one-dimensional effective interaction strength as a function of the dimensionless lattice depth for the two different approaches defined by Eqs.~\eqref{tightBinding} and \eqref{interactionWannier}. Using the methods of Ref.~\cite{krutitsky16} the value for the effective interaction strength in the pure 3D regime is found to be $\lim_{s\rightarrow0}g_{\mathrm{eff}}^{\mathrm{1D}}a^2=4g/9$. For deep lattices the relative error between the two methods decreases.

\section{Critical Chemical Potential}
\label{sec5}

\begin{figure}
\begin{flushleft}
\includegraphics[width=.98\columnwidth]{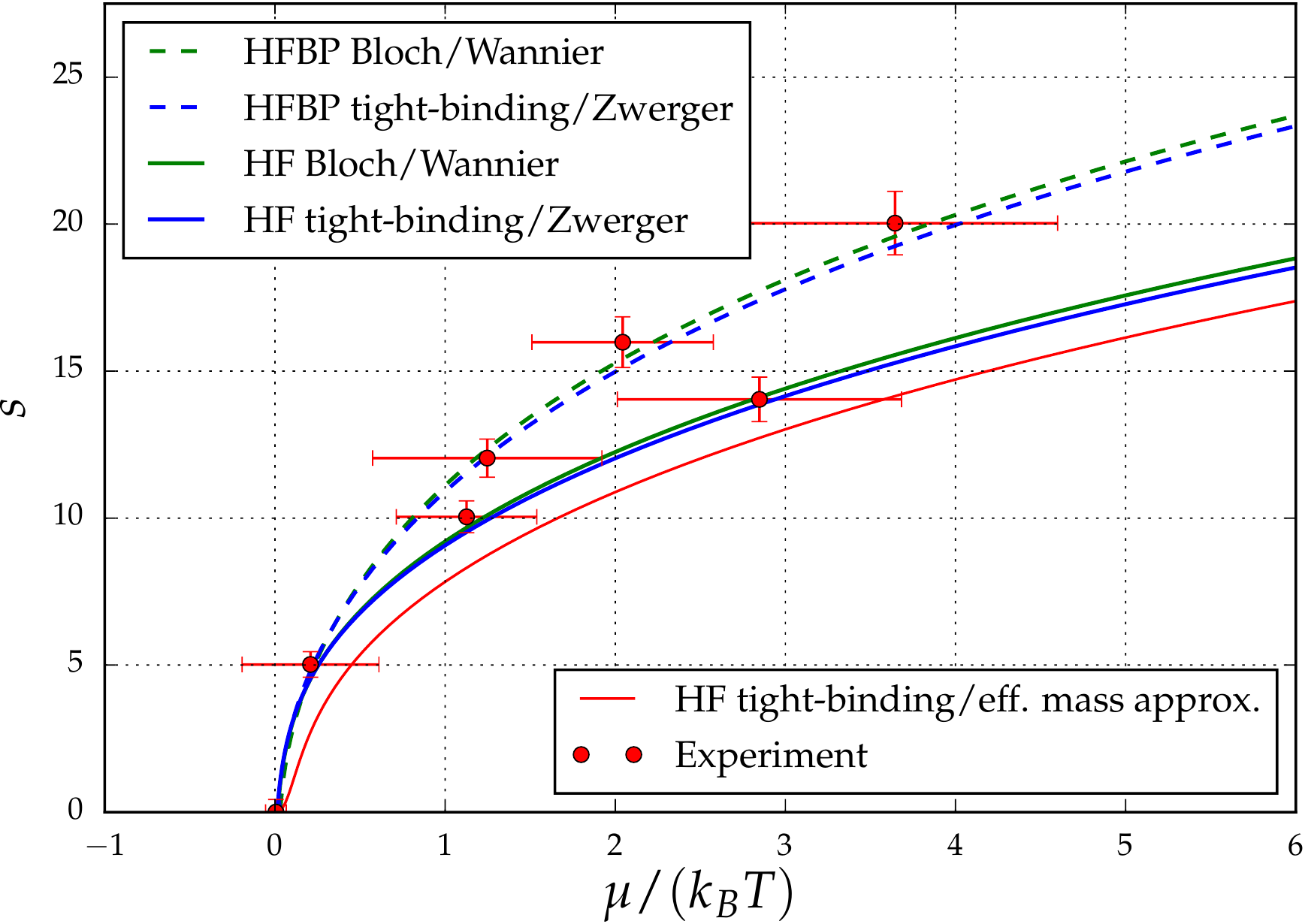}
\caption{Phase diagram in $s$-$\mu$-plane: red data points are reproduced from Ref.~\cite{vogler14} and describe phase boundary between decoupled 1D tubes and 3D condensate, red lower curve is HF result from Ref.~\cite{vogler14}, solid lines represent HF, and dashed lines HFBP results.}
\label{fig:chemicalPotential}
\end{flushleft}
\end{figure}

Now we determine within the HFBP theory the critical chemical potential $\mu_c$ as a function of the lattice depth $s$ as it was measured experimentally via LDA, as well as calculated within a HF treatment of Ref.~\cite{vogler14}. 

In the thermodynamic limit the direct numerical integration in Cartesian coordinates of Eq.~\eqref{HFBPdensity} would lead to a divergence at $\bm{k}=\bm{0}$. However, in elliptical coordinates this divergence is avoided due to a factor from the integration measure. Therefore, we perform the integration by cutting a small ellipsoid around the origin, which can then be calculated analytically, whereas the remaining integration volume of the Brillouin zone is computed numerically. For better convergence we use a quadratically aligned sampling of integration points. Thus, the sampling is quite dense around the origin which makes the result robust against the choice of the ellipsoid.

The critical chemical potential describes the phase boundary between decoupled 1D tubes and the 3D condensate as depicted in Fig.~\ref{fig:chemicalPotential}. Therein, we present the HF treatment, as well as the HFBP results and compare them with the experimental data of Ref.~\cite{vogler14}. The red lower curve corresponds to the HF result in the tight-binding approximation for the effective interaction strength and the effective mass approximation for the hopping energy as described above. It coincides with the result presented in Ref.~\cite{vogler14}. Note however, that in Ref.~\cite{vogler14} the critical chemical potential is erroneously taken to be $\mu_c=gn$, although a proper HF treatment yields $\mu_c=2gn$. Therefore, the data of Ref.~\cite{vogler14} have been multiplied {ad hoc} by a factor 2 for the sake of comparison. The remaining solid lines are HF results with improved methods for both the effective interaction strength and the hopping energy. Since the upper solid blue and green HF curves differ by much less than the size of the error bars, we conclude that the largest error source for the underestimation of the red curve is the effective mass approximation. However, since Eq.~\eqref{zwerger} is known to reproduce imprecise values for shallow lattice depths, we understand this rather good result,   represented by the upper solid blue curve, as a canceling of errors of the hopping energy \eqref{zwerger} and the tight-binding approximation \eqref{tightBinding}. In contrast to this the green curves stem from hopping energies and interaction strengths, which are computed by numerically exact Wannier functions. The dashed curves represent the HFBP results which are given by Eq.~\eqref{GPequation}. We observe that they are in much better agreement with the experimental data over the full range of the dimensionless lattice depth $s$ than the HF results. But note that all curves coincide in the regime of shallow lattices, where the system close to 3D. As a consequence, we ascribe the better agreement of the HFBP theory with the experimental data in the low-dimensional regime to the enhanced role of quantum fluctuations which are neglected in the HF treatment.

\section{Critical Temperature}
\label{sec6}

\begin{figure}
\centering
\includegraphics[width=\columnwidth]{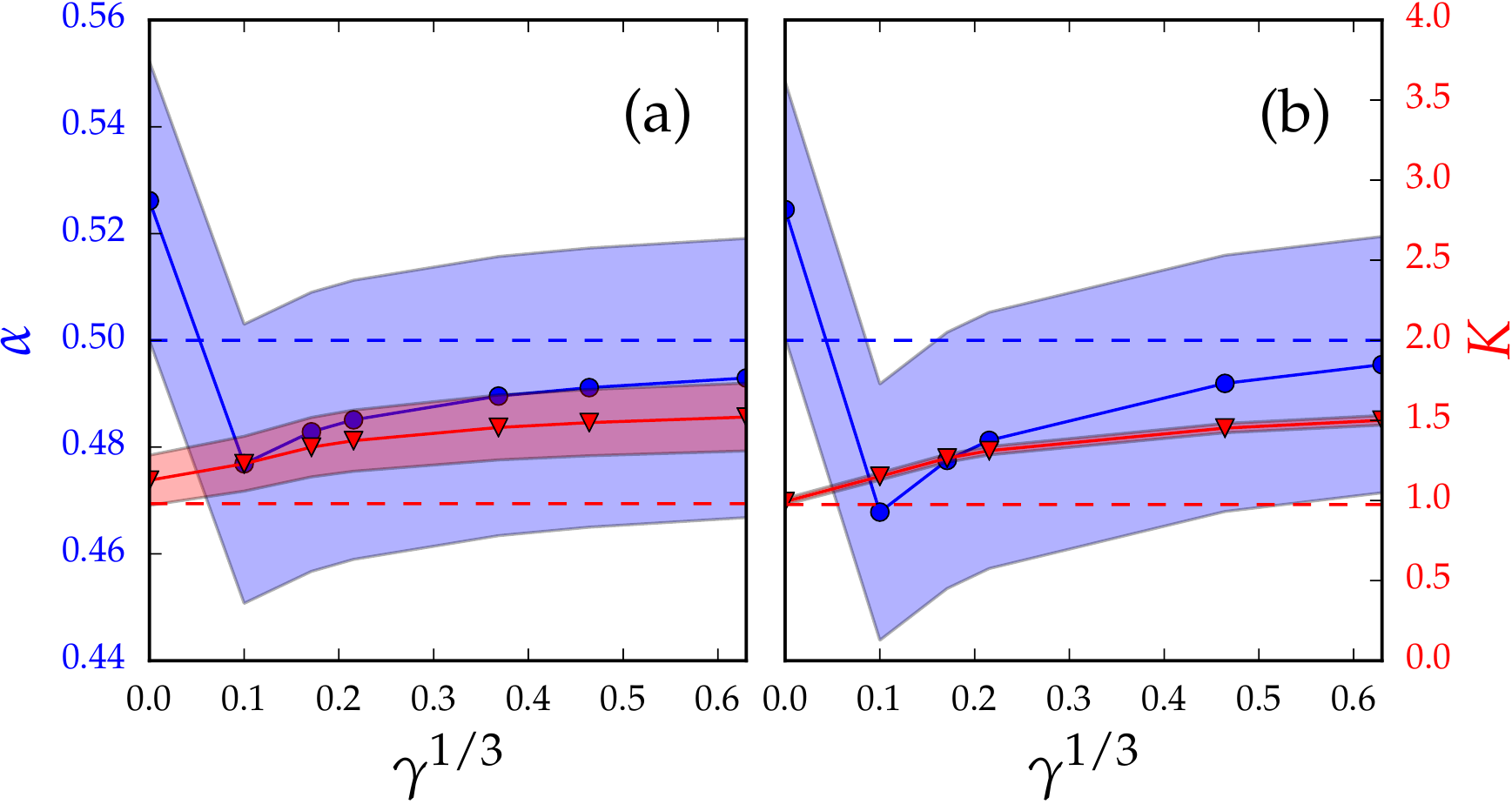}
\caption{Power-law exponent $\alpha$ (circles) and prefactor $K$ (triangles) as functions of gas parameter $\gamma^{1/3}=n^{1/3}a_s$ for (a) {hybrid model} and (b) {pure lattice model}. Dashed lines represent corresponding values for non-interacting case, upper ones for $\alpha$, lower ones for $K$.}
\label{fig:powerLaw}
\end{figure}

We now study the power-law behavior of the critical temperature of the {hybrid model} at the dimensional phase transition, which has not yet been measured experimentally. As we have for no interactions the result  \eqref{HFmodelA}, we assume for a finite two-particle interaction near the transition a general power-law

\begin{equation}
\frac{k_BT_c}{E_r}= K \left(\frac{J}{E_r}\right)^\alpha,
\end{equation}
with $\alpha$ and $K$ being the exponent and the prefactor, respectively. In Fig.~\ref{fig:powerLaw}\,(a) we present the numerical results of the HFBP theory for finite interaction strengths.
Here, we set the density to $n=13.3$\,nm$^{-3}$ and the lattice constant to $a=387$\,nm, which are taken from the experiment of Ref.~\cite{vogler14}. The shaded areas in Fig.~\ref{fig:powerLaw} represent error estimates due to fitting errors. They correspond to the difference between the numerically determined value of the non-interacting case and the value, which is already known from Eq.~\eqref{HFmodelA}, and amounts to approximately 5\% for the exponent $\alpha$ and 13\% for the prefactor $K$. We observe that within our precision the exponent does not change with finite interactions, however the prefactor does. Thus, we conclude that the exponent exhibits a surprising robustness with respect to the interaction strength.

Due to this robustness the question arises how far the exponent of 1/2 is universal. In order to analyze this at least exemplarily, we apply the same study to a different model of the same 1D-3D phase transition. This model is a {pure lattice model} and is investigated with high-precision quantum Monte Carlo data from Ref.~\cite{morath16}. In contrast to the {hybrid model} \eqref{hybridModel}, it is characterized by
\begin{equation}
J_x=J_y=J\neq J_z,\quad 
a_x=a_y=a_z=a.
\end{equation}
Here the dimensional transition is induced through the tunable ratio of the hopping energies $J/J_z$. If the transverse hopping energy is much smaller than the longitudinal one, i.e., $J_z\gg J$, the atoms within the lattice are only allowed to hop in longitudinal direction, which represents the 1D regime. However, for similar hopping energies $J_z\approx J$ the system corresponds to a 3D lattice.

In the non-interacting case the HF density \eqref{HFdensity} in the thermodynamic limit can be integrated analytically and reads 
\begin{equation}
\begin{split}
n=n_0+\frac{1}{a^3}\sum_{m=1}^\infty
&e^{m\beta(\mu-2gn)}e^{-4m\beta J}I_0^2(2m\beta J)\\
&\times e^{-2m\beta J_z}I_0(2m\beta J_z).
\end{split}
\end{equation}
Correspondingly, the power-law of the critical temperature follows as \cite{morath16}
\begin{equation}
\frac{k_BT_c}{E_r}=4na^3\sqrt{\frac{J_z}{E_r}}\sqrt{\frac{J}{E_r}}.
\label{HFmodelB}
\end{equation}
Hence, from Eqs.~\eqref{HFmodelA} and \eqref{HFmodelB} we read off that both models turn out to have the same power-law exponent for the increase of the critical temperature in the 1D-3D-transition close to 1D for the non-interacting case. 

Based on these findings, we determine with the HFBP theory also the power-law parameters for the weakly interacting gas within the {pure lattice model}. To this end we set the longitudinal hopping energy $J_z=0.1\,E_r$, which corresponds to a shallow lattice according to Fig.~\ref{fig:hopping}. As depicted in Fig.~\ref{fig:powerLaw}\,(b), we find the same robustness of the exponent as in the {hybrid model} with an error of 5\% for the exponent $\alpha$ and 3\% for the prefactor $K$.

\section{Conclusion and Outlook}
\label{sec8}
We conclude that the HFBP theory turns out to be in a very good agreement with the experimental data for the critical chemical potential of the {hybrid model}. Since the HFBP theory includes quantum fluctuations, it represents a crucial improvement to the HF theory when studying quantum systems in low dimensions at finite temperature.
Furthermore,  we investigated two different models which induce the dimensional phase transition between decoupled 1D tubes and the 3D BEC within the formalism of the HFBP theory including numerically exact hopping energies and interaction strengths. The {hybrid model} and the {pure lattice model} are found to exhibit within our accuracy the same power-law exponent of 1/2 for the critical temperature during the dimensional phase transition. This exponent is robust against change of the finite interaction strength, which seems to put both systems into the same universality class. 

Dimensional phase transitions examine the behavior of observables as function of a control parameter of the effective dimension. In our case this is the transverse hopping energy. The idea of dimensional phase transitions can straight-forwardly be generalized in order to study, e.g., the 2D-3D-transition, where the critical temperature increases instead of Eq.~\eqref{HFmodelB} with a logarithm-like behavior. Furthermore thermodynamic quantities such as the heat capacity or trapped systems can be investigated as well. In the latter case the Mermin-Wagner-Hohenberg theorem does not hold anymore since they are not translationally invariant. 

Also in the case of fermions the dimensional transition is of significant interest. For instance, a gas of spin imbalanced fermions is a candidate for the Fulde-Ferrell-Larkin-Ovchinnikov phase (FFLO) \cite{fulde64,larkin65}. Fundamental work for the exploration of this exotic phase has been done theoretically  for the 1D-3D-transition in the framework of dynamical mean-field theory \cite{heikkinen14} and static mean-field theory \cite{dutta15} by investigating the stability of the FFLO phase. Furthermore, a recent experiment \cite{revelle16},  showed an interesting spatial ordering of the polarized and non-polarized phases and its inversion during the 1D-3D-transition.

\section*{Acknowledgment}
The authors gratefully thank Antun Bala\v{z}, Sebastian Eggert, Denis Morath, Herwig Ott, and Dominik Stra\ss el for fruitful discussions. 
Furthermore, we acknowledge support of binational projects funded by the German Academic Exchange Service (DAAD) and the Brazilian Federal Agency for Support and Evaluation of Graduate Education (CAPES), as well as the Ministry of Education, Science, and Technological Development of the Republic of Serbia. Finally, we thank the German Research Foundation (DFG) via the Collaborative Research Centers SFB/TR 49 and SFB/TR 185 for financial support.

\bibliographystyle{apsrev4-1}
\bibliography{bibliography}
\end{document}